\documentstyle{mn}
\newif\ifAMStwofonts
\def\overleftrightarrow{\mathpalette\overleftrightarrow@}
\def\overleftrightarrow@#1#2{\vbox{\ialign{##\crcr\leftrightarrowfill@#1\crcr
 \noalign{\kern-\ex@\nointerlineskip}$\m@th\hfil#1#2\hfil$\crcr}}}
\input psfig.tex
\input tcilatex  

\ifoldfss
  \ifCUPmtlplainloaded \else
    \NewTextAlphabet{textbfit} {cmbxti10} {}
    \NewTextAlphabet{textbfss} {cmssbx10} {}
    \NewMathAlphabet{mathbfit} {cmbxti10} {} 
    \NewMathAlphabet{mathbfss} {cmssbx10} {} 
  \fi
  \ifAMStwofonts
    \ifCUPmtlplainloaded \else
      \NewSymbolFont{upmath} {eurm10}
      \NewSymbolFont{AMSa} {msam10}
      \NewMathSymbol{\upi}     {0}{upmath}{19}
      \NewMathSymbol{\umu}     {0}{upmath}{16}
      \NewMathSymbol{\upartial}{0}{upmath}{40}
      \NewMathSymbol{\leqslant}{3}{AMSa}{36}
      \NewMathSymbol{\geqslant}{3}{AMSa}{3E}

    \fi
  \fi
\fi 
\ifnfssone
  \newmathalphabet{\mathit}
  \addtoversion{normal}{\mathit}{cmr}{m}{it}
  \addtoversion{bold}{\mathit}{cmr}{bx}{it}
  \newmathalphabet{\mathbfit} 
  \addtoversion{normal}{\mathbfit}{cmr}{bx}{it}
  \addtoversion{bold}{\mathbfit}{cmr}{bx}{it}
  \newmathalphabet{\mathbfss} 
  \addtoversion{normal}{\mathbfss}{cmss}{bx}{n}
  \addtoversion{bold}{\mathbfss}{cmss}{bx}{n}
  \ifAMStwofonts
    \ifCUPmtlplainloaded \else
      \UseAMStwoboldmath
      \makeatletter
      \new@mathgroup\upmath@group
      \define@mathgroup\mv@normal\upmath@group{eur}{m}{n}
      \define@mathgroup\mv@bold\upmath@group{eur}{b}{n}
      \edef\UPM{\hexnumber\upmath@group}
      \new@mathgroup\amsa@group
      \define@mathgroup\mv@normal\amsa@group{msa}{m}{n}
      \define@mathgroup\mv@bold\amsa@group{msa}{m}{n}
      \edef\AMSa{\hexnumber\amsa@group}
      \makeatother
      \mathchardef\upi="0\UPM19
      \mathchardef\umu="0\UPM16
      \mathchardef\upartial="0\UPM40
      \mathchardef\leqslant="3\AMSa36
      \mathchardef\geqslant="3\AMSa3E
    \fi
  \fi
\fi 

\ifnfsstwo
  \DeclareMathAlphabet{\mathbfit}{OT1}{cmr}{bx}{it}
  \SetMathAlphabet\mathbfit{bold}{OT1}{cmr}{bx}{it}
  \DeclareMathAlphabet{\mathbfss}{OT1}{cmss}{bx}{n}
  \SetMathAlphabet\mathbfss{bold}{OT1}{cmss}{bx}{n}
  \ifAMStwofonts
    \ifCUPmtlplainloaded \else
      \DeclareSymbolFont{UPM}{U}{eur}{m}{n}
      \SetSymbolFont{UPM}{bold}{U}{eur}{b}{n}
      \DeclareSymbolFont{AMSa}{U}{msa}{m}{n}
      \DeclareMathSymbol{\upi}{0}{UPM}{"19}
      \DeclareMathSymbol{\umu}{0}{UPM}{"16}
      \DeclareMathSymbol{\upartial}{0}{UPM}{"40}
      \DeclareMathSymbol{\leqslant}{3}{AMSa}{"36}
      \DeclareMathSymbol{\geqslant}{3}{AMSa}{"3E}
    \fi
  \fi
\fi 

\ifCUPmtlplainloaded \else
  \ifAMStwofonts \else 
    \def\upi{\pi}
    \def\umu{\mu}
    \def\upartial{\partial}
  \fi
\fi

\title{Smoothed Particle Hydrodynamic Simulations of Viscous Accretion Discs Around Black Holes$^\dag$}
\author[Giuseppe Lanzafame, D. Molteni and S.K. Chakrabarti]
{Giuseppe Lanzafame$^1$, D. Molteni$^2$ and Sandip K. Chakrabarti$^3$ \\
$^1$  Osservatorio Astronomico, University of Catania, Catania, ITALY\\
$^2$ Istituto di Fisica, Via Archirafi 36, 90123 Palermo, ITALY\\
$^3$ S.N. Bose National Centre for Basic Sciences, JD-Block, Sector-III, 
Salt Lake, 700091, INDIA\\
$^\dag$ MNRAS (Submitted); SNBNCBS preprint No: SNB/ASTRO/3-97}

\date{Accepted ....  Received ... ; in original form 21st June, 1996}

\pubyear{1997}

\begin{document}

\maketitle

\begin{abstract}
Viscous Keplerian discs become sub-Keplerian close to a black hole since 
they pass through sonic points before entering into it. We study the time 
evolution of polytropic viscous accretion discs (both in one and two
dimensional flows) using Smoothed Particle Hydrodynamics. We
discover that for a large region of the parameter space, when the flow
viscosity parameter is less than a critical value, standing shock waves 
are formed. If the viscosity is very high then the shock disappears.
In the intermediate viscosity the disc oscillates very significantly
in viscous time-scale. Our simulations indicate that these centrifugally 
supported high density region close to a black hole plays an active role 
in the flow dynamics, and consequently, the radiation dynamics.
\end{abstract}


\newpage

\noindent {\bf{\large  1. INTRODUCTION}}

One of the convincing ways to understand the nature of accretion processes
on black holes is to use numerical techniques. This is because the steady 
solutions of analytical models are too simplistic and they usually concentrate 
on steady state solutions. The works present in the literature, which study the 
possibility of sub-Keplerian accretion flows are mainly carried out for inviscid flows 
(Hawley, Smarr \& Wilson, 1984; Molteni, Lanzafame \& Chakrabarti, 1994; [hereafter MLC94]
Nobuto \& Hanawa, 1994; Molteni, Ryu \& Chakrabarti, 1996 [hereafter MRC96];
Ryu, Chakrabarti \& Molteni, 1997 [hereafter RCM97])
although some simulations of viscous {\it Keplerian} discs are also present 
(Eggum, Coroniti \& Katz, 1988). In Chakrabarti \& Molteni (1995, hereafter CM95; 
and references therein), the time evolution of {\it one dimensional isothermal}
viscous transonic flows (VTFs) has been studied. It was particularly noted that in some 
region of the parameter space, when the viscosity is smaller,
the flow developed centrifugal pressure supported shock waves as is expected
from the steady solutions (Chakrabarti, 1990; hereafter C90). As viscosity is increased,
the angular momentum transport rate in the post-shock region is 
enhanced compared to the rate in the pre-shock flow and as a result the shock 
propagates outwards, and the flow becomes sub-sonic. Further out, the
flow can join with a Keplerian accretion disc. 

In the present paper, we study the behaviour of the sub-Keplerian VTFs
when the flow is neither isothermal nor restricted only to the equatorial plane
as in CM95. We examine the behaviour of the more general polytropic 
thick, viscous accretion discs. We find that even in two dimensional thick discs, shocks
form and the steady shock location increases with viscosity as in one 
dimensional study of CM95. We also observe, to our surprise, that beyond a critical
viscosity, when the steady shock is not expected, the flow forms
an unsteady shock which periodically evacuates the disc. 
Beyond another critical viscosity (keeping other parameters unchanged), 
where the flow passes only through the inner sonic point, 
the shock disappears and only the smooth sub-Keplerian disc 
(originating from a Keplerian disc) remains. 
The presence oscillating solutions are clear indications of 
transitions from one topology of solutions to another.
Presence of both steady and oscillating solutions 
is significant in view of the fact that spectral observations of black hole
candidates  do show quasi-periodic oscillations in between various 
stationary states. Simply put, two critical viscosities in the above context
come into being for the following reasons: Beyond a critical viscosity,
the shock conditions cease to be fulfilled even when the flow
has two saddle type sonic points. The flow passes through both,
using unsteady shocks.
Beyond yet another, and higher value of critical viscosity, the outer saddle type 
sonic point cease to exist altogether and flow smoothly passes through the
inner sonic point. The general behaviour of inviscid 
and viscous accretion flows and how these critical viscosities can be identified,
are discussed in C90 and Chakrabarti (1996, hereafter C96) respectively.

The organization of the present paper is the following: In the next 
Section, we briefly discuss the theory of viscous transonic discs.
In \S 3, we present the simulation results in one dimensional
viscous discs. In \S 4, we present the results in two dimensional
viscous discs. In \S 5, we discuss astrophysical importance of our 
solutions and present concluding remarks.

\noindent {\bf{\large 2. MODEL EQUATIONS}}

We model the nonrotating central compact object 
using Paczy\'nski \& Wiita (1980) potential. 
The inclusion of viscosity in the form of Shakura-Sunyaev prescription 
(or, other viscosity prescriptions as in Chakrabarti \& Molteni, 
1995) implies that the specific angular momentum $\lambda$ is not 
constant everywhere, rather it is efficiently transported outward
at the rate determined by the magnitude of viscosity.

We measure all distances, velocities and timescales in units of $2GM/c^{2}$, 
$c$ and $2GM/c^{3}$ respectively. Below, we provide the Lagrangean 
formulae for the two-dimensional fluid dynamics equations for 
SPH in cylindrical coordinates. The mass conservation equation is:
$$
\frac{D\rho }{Dt}=-\rho \nabla {\bf v} 
$$
(here, $\frac D{Dt}$ is the comoving derivative)

The momentum equation is:
$$
\frac{D{\bf v}}{Dt}=-\frac 1\rho \nabla P+{\bf g}+\frac{v_\phi ^2}r\widehat{r%
}-\frac{v_\phi v_r}r\widehat{\phi }+\frac 1\rho \nabla \overleftrightarrow{%
{\bf \tau }} 
$$
Where $\widehat{r}$ is the radial direction vector, $\widehat{\phi }$ is the
tangential vector and,
$$
{\bf g}=-\frac 1{2\left( R-1\right) ^2}\frac{{\bf R}}R,
$$
$$
g_r=-\frac 1{2\left( R-1\right) ^2}\frac rR,
$$
$$
g_z=-\frac 1{2\left( R-1\right) ^2}\frac zR, 
$$
here, $R=r \widehat{r}+z \widehat{z}$ 
$$
R=\sqrt{r^2+z^2} 
$$
The radial momentum equation becomes,
$$
\frac{Dv_r}{Dt}=-\frac 1\rho \frac{\partial P}{\partial r}+g_r+\frac{v_\phi
^2}r  .
$$
The vertical momentum equation becomes,
$$
\frac{Dv_z}{Dt}=-\frac 1\rho \frac{\partial P}{\partial z}+g_z ,
$$
The tangential momentum equation becomes,
$$
\frac{Dv_\phi }{Dt}=-\frac{v_\phi v_r}r+\frac 1\rho \left[ \frac 1{r^2}\frac
\partial {\partial r}\left( r^2\tau _{r\phi }\right) \right] 
$$
where,
$$
\tau _{r\phi }=\mu r\frac{\partial \Omega }{\partial r}, 
$$
$$
\Omega =\frac{v_\phi }r .
$$
For the Shakura Sunyaev turbulent viscosity we have used,
$$
\mu =\alpha \cdot \rho \cdot c_{sound}\cdot Z_{disc} 
$$
where the vertical thickness Z$_{disc}$ is estimated from the
assumption of the vertical equilibrium condition:
$$
Z_{disc}=\frac 2\gamma \cdot c_{sound}\cdot r\cdot (r-1)^2.
$$

One may use the conventional energy equation as,
$$
\frac{D\epsilon }{Dt}=-\frac P\rho \nabla {\bf v+}\frac \Phi \rho =-\frac
P\rho \frac 1r\frac \partial {\partial r}\left( rv_r\right) +\frac \mu \rho
\left[ r\frac{\partial \Omega }{\partial r}\right] ^2 ,
$$
where, $\epsilon$ is the specific thermal energy, and $\Phi {\bf \ }$ 
is the usual viscous dissipation term. However, it turns out that a 
better accuracy is achieved if the sum total of specific kinetic and thermal
energies are used instead. This is because this quantity is exactly
conserved (since it can be put in a symmetric form so that pair of particles
exchange equal amounts of energy, e.g. Monaghan 1985). So we prefer
to use the following version of the energy equation:

$$
\frac D{Dt}\left( \epsilon +\frac 12{\bf v}^2\right) =-\frac P\rho \nabla
{\bf v+v\cdot }\left( \frac{D{\bf v}}{Dt}\right) +\frac 1\rho \nabla \left(
\bf \overleftrightarrow{ {\bf \tau } } {\bf :v} \right)
$$

with 

$$
\left( \frac{D{\bf v}}{Dt}\right) =-\frac 1\rho \nabla P+{\bf g} 
$$
where $P=\left( \gamma -1\right) \rho \epsilon $ is the equation of state of
ideal gas, {\bf g} is the gravitational acceleration. ${\bf 
\overleftrightarrow{{\bf \tau }}:v}$ is the vector resulting from the
contraction of the stress tensor with the velocity vector.
We include only $\tau _{r\phi }$ (namely, the $r\phi$ component)
since it is the dominant contributor to the viscous stress.
We choose for simplicity the pseudo-Newtonian potential proposed by
Paczy\'nski \& Wiita (1980), where $\Phi(r, \theta)= 1/2(r-1)$.
A complete steady solution requires the equations of energy, angular momentum 
and mass conservation supplied by transonic conditions at the critical 
points and the Rankine-Hugoniot conditions at the shock.

Since a black hole accretion is necessarily supersonic at the
horizon (Chakrabarti, 1990), and since the flows are sub-Keplerian
at the sonic point even for generalized accretion models (C96), 
a flow must deviate from a Keplerian disc much before it enters into a hole.
We verify this by performing numerous numerical simulations. If the
viscosity is smaller than a critical value, and if the shock conditions
are satisfied, then the shocks form, otherwise the shock becomes
unstable and propagate outwards to make a subsonic disc which joins
with a Keplerian disc. The reason for the shock to propagate outwards
is that the rate of angular momentum transport in the pre and post-shock flows 
become completely different. In particular, the transport rate is higher
in the subsonic post-shock region. Even when the shock is absent, the
centrifugal barrier supported region close to the black hole
behave very similar to the post-shock region, and hence the
emission properties, of this shock-free disc also becomes similar (Chakrabarti, 1997).

In this context, the difference between the approach to obtain the  steady 
solutions (as in C90, or in C96) and time dependent
solutions is worth noting. While obtaining a steady solution, since the
flow must pass through a sonic point, it is customary to choose 
it's location and the specific angular momentum of the flow at this point.
The rest of the properties of the flow (such as the location where the
flow deviates from a Keplerian disc, it's injection velocities etc.)
are determined as eigenvalues. In the time-dependent studies, on the other hand,
we fix the injected parameters (such as the specific energy
and angular momentum) while the location of the sonic point and the angular
momentum at the sonic point are determined automatically as eigenvalues. That these
two approaches produce the same result (whenever steady state solutions
exist) has been verified repeatedly (See, first paragraph of the Introduction
for reference.). In the present paper also, we shall fix the parameters 
of the injecting matter at the outer boundary.

The numerical simulations we perform are done using Smoothed Particle 
Hydrodynamics (SPH). The code has been tested several times against 
theoretical works (e.g., CM93, CM95; Sponholz \& Molteni, 1994; 
MRC96; RCM97) and is found to be sufficiently accurate as far as the
angular momentum conservation is concerned. In the next section we present 
results in one dimensions and in \S 4, we present results in 2 dimensional disc.

\noindent {\bf{\large 2. SIMULATION RESULTS IN ONE DIMENSIONS}}

Figure 1 shows stationary shock locations as a function of 
Shakura-Sunyaev viscosity parameter $\alpha$ in thin, rotating,
accreting flows. The viscosity coefficient $\nu$ is related to the
viscosity parameter $\alpha$ by  (see, Shakura \& Sunyaev, 1973)
$\mu =\alpha \rho a Z_{disc}$.
In upper left panel, we plot density $\rho (r) $, in upper 
right panel, we plot radial velocity $v_r (r)$, in lower left panel, 
we plot $\lambda (r)$ and the in lower right panel, we plot Mach number
$M_x (r)$ [same as $ M_r (r) $ in axial direction]. Matter with sound 
velocity $a=0.12272$, injection velocity $v_{in}=0.12319$ and 
specific angular momentum $\lambda=1.8$ is injected at $X_{in}=26.15$. 
Each result has been obtained starting with an inviscid 
flow $\alpha=0.0$ and then increasing $\alpha$ up to $4.6 \cdot 10^{-4}$. 
In {\it each} case, standing shock solution was obtained. 
Note, that in this particular case, the shock disappears 
around $\alpha=4.6\times 10^{-4}$ which is the critical $\alpha$ here. 
The shock location shifts outwards as described in C90 and C96
with viscosity, and the angular momentum becomes more and more 
Keplerian. The results are indeed similar to Chakrabarti \& Molteni 
(1995) solution of viscous isothermal flows. As viscosity is increased, 
the width of the shock also increases as it should be. For a different
choice of injected parameters, the critical $\alpha$ could be very high
as well (C96).

In Fig. 2, we use the injected angular momentum to be 
same as the marginally stable value $\lambda=1.8363$ at the
outer boundary located at $X_{in}=50$. Injection velocity of matter
and the sound speed are $0.157$ and $0.0738$ respectively. The stationary
shock locations are seen in all the viscosity parameter values.
The critical viscosity in this particular case is much higher,
and therefore the shock continues to persist for the range of
$\alpha$ used. The angular momentum distribution becomes closer
to Keplerian much faster for higher values of viscosity parameters. 
However, the flow remains {\it sub-Keplerian} throughout. By suitably
adjusting the injection parameters at the outer boundary one obtains the
solutions reaching a Keplerian disc. This requires very large number of 
smoothed particles.

\noindent {\bf{\large 3. LOCATION OF STANDING SHOCKS IN THICK VISCOUS 
ACCRETION DISCS}}

In this Section, we describe the formation of standing shocks in 
rotating, {\it thick}, axisymmetric, accretion discs onto a black 
hole. We introduce viscosity to the inviscid solutions
described in detail by Molteni, Lanzafame \& Chakrabarti, 1994
(hereafter referred to as MLC94). In MLC94, the entire parameter
space spanned by the specific energy {\large $\epsilon$} 
and specific angular momentum $\lambda$ was depicted which
shows the boundary between shock and no-shock solutions in a flow 
in vertical equilibrium. It was observed that the theoretical 
predictions based on flow models in vertical equilibrium regarding 
the shock locations are generally in good agreement with the
simulation results when appropriate effects of turbulence is
taken care of.

As in the one-dimensional simulations presented in the previous section,
we introduce viscosity gradually also in the two dimensional simulations,
and in {\it each} case, we continue simulations till
a stationary solution is reached. Secondly, unlike in MLC94,
where the simulations were carried out with matter injected
only in one quadrant, here we inject matter in two quadrants --
both the upper and the lower side of the equatorial plane.
As in MLC94, the accretion rate $\dot{{\it M}}$ being an 
eigenvalue of the problem, it is fixed by the choice of the input 
parameters ({\large $\epsilon$}, $\lambda$). Therefore, matter density is 
chosen to be equal to unity. Since the accretion rate is an eigenvalue
of the problem, it is automatically adjusted once the specific angular
momentum and the specific energy are fixed at the outer boundary. This 
is true when a steady state is reached. When a steady state is not 
reached, such considerations do not apply.

Figure 3 shows the XZ location of the SPH particles of the inviscid 
accretion disc model. Particles with specific angular momentum 
$\lambda=1.6$ and energy {\large $\epsilon$} $=0.001955219$ (which 
corresponds to injected radial velocity and the sound speed as $v_{in} 
=0.121122$ and $a_{in}=0.058989$ at $X_{in}=30$ respectively) 
are injected from the outer edge of the disc at $X=30$. 
The results in the upper left panel is for inviscid 
flow. It is shown at $t=2900$ when the flow has achieved 
a steady state. The total number of particles, whose size is $h= 0.3$, 
is $6146$. A stationary shock is formed around $X \sim 5.2$ 
and weaker oblique shocks are also seen (e.g., MLC94). At the 
two wings of the hot post-shock flow  (basically inside the
centrifugal barrier and the funnel wall, see, MRC96) a subsonic 
cooler outflow is present, symmetrically both up and down which 
carry away of the order of $5\%$ of the total mass of the disc. No 
turbulent motions are evident in this model in the narrow post-shock 
subsonic flow. Subsequent to the shock, the hot 
flow becomes supersonic again before entering inside the black hole.

In the other three panels we show the effect of the introduction
of viscosity in the flow. Clearly, as in the one-dimensional
case, higher viscosity causes higher differential angular momentum transport 
between the pre- and the post-shock solutions and as a result the
shock is drifted away in the radial direction. The value of the
viscosity parameter $\alpha$ is written on each panel. In the increasing
order of viscosity, the shocks locations are $X_{s}=6.0, \ 7.5,$ and $ 10.5$
respectively. Another point of interest: as viscosity is raised, the 
amount of outflowing matter in the wind is decreased. This is because
of weakening of the centrifugal barrier of the in-going matter. Lower viscosity
causes matter to bounce from the barrier and fly away as winds. In higher 
viscosity, higher turbulences are also seen to form in the post-shock flow.
This is because more matter from higher elevation falls on the equatorial
plane and convert their potential energy to turbulent energy.

It is now widely believed (see, e.g., Chakrabarti, 1997 and references 
therein) that the observed X-rays in black hole candidates are produced and 
reprocessed  in the centrifugal barrier supported regions very close to 
the black holes. The soft photons released from the pre-shock Keplerian 
disc is reprocessed in the centrifugal pressure supported region (with or without shock)
to become hard X-rays. Occasionally, radiations show quasi-periodic
oscillations. Molteni, Sponholz \& Chakrabarti (1996), and  Ryu, Chakrabarti \& Molteni
(1997) suggest that these so-called QPOs observed in the compact 
objects could be due to the oscillations of the `boundary layers', 
i.e., the centrifugally supported denser region. These oscillations form 
for those flow parameters where the theoretical analysis does 
not predict the formation of steady shocks, even though two
sonic points exist. However, so far, no viscous flow was simulated. 
In Fig. 4a we show the first simulations of viscous flow shocks oscillate 
periodically. In the upper-left, upper-right and lower-left panels roughly
half of the cycle is shown where the shock location decreases
monotonically. In the lower-right panel, the shock drifted again
outward. The parameters are same as in Fig. 3, but the viscosity
is higher: $\alpha=2 \times 10^{-3}$. The time period is around 
$T_{QPO}=4000 $ in units of $2GM/c^2$. Thus, in a stellar black hole
of mass $10 M_\odot$, the period would be around $0.4$s, whereas
in a supermassive black hole of mass $10^7 M_\odot$, the period would be, 
$10^6$s or a few days. These time scales are comparable with the
time scale of observed QPOs in black hole candidates. With a different
disc input parameters (e.g., for different accretion rate or 
viscosity of the original disc) the periods will vary.
Clearer evidence of the oscillation of the shock is shown in Fig. 4b, where the matter
close to the equatorial place is collected and their Mach numbers 
are plotted. The behaviour of the shock on the equatorial plane
is similar to what is seen in the one-dimensional simulations of
Molteni, Sponholz \& Chakrabarti, 1996.

\noindent {\small {\bf Fig. 4b:}
Mach number of the particles close to the equatorial plane is
plotted against radial distance at different phases of oscillation
for the case shown in Fig. 4a}                 

Figure 5 shows the variation of the number of simulation particles
as a function of time in the case shown in Fig. 4a. Here the total
number $N$ and the number of sub-sonic particles $N_{sub}$ (presumably,
participating in the Compton reprocessing of the soft-photons from the
Keplerian disc) are shown. The amplitude modulation is significant: $50$ 
percent variation in total particle number, and more than $700$ percent variation 
in the sub-sonic particle number. We believe that this is important: the 
observed significant ($10-100$ percent) variations in QPO cannot 
be explained away by any means other than such a dynamical variation of the
X-ray emitting region. 

\noindent {\bf\large 4. CONCLUDING REMARKS}

In the present paper, we have numerically studied the behaviour of viscous, transonic
flows. There are several important conclusions: We show that the location of
the centrifugally driven standing shock wave drifts away from the black hole
and ultimately the shock disappears. This is valid both for one and two dimensional
axisymmetric discs originating from Keplerian flows. This result 
generalizes the earlier study of Chakrabarti
\& Molteni (1995) for isothermal flows done in one spatial dimensions. Second
most important conclusion is that the viscous flows also show shock
oscillations in regions of the parameter space where the steady shock
is not possible even when two saddle type sonic points are present. This generalizes earlier
work of Ryu, Chakrabarti \& Molteni (1997) where non-viscous flow was studied
in two dimensions, and that of Molteni, Sponholz \& Chakrabarti (1996) where 
non-viscous flow was studied in presence of bremsstrahlung cooling. We believe 
that both these findings are very important in explaining the observed X-ray 
radiations from the black hole candidates. The drifting shock
solutions present the insight of how a Keplerian disc is actually formed
out of an original sub-Keplerian flow and how the centrifugal pressure supported region
in the sub-Keplerian region around the black hole may act as the so-called Compton 
cloud. The dynamical oscillation of the centrifugal
pressure supported shock wave (which, for the first time is shown here
to be present even in a viscous transonic disc) produces right frequency and 
the amplitude of modulation to be taken a serious explanation of 
the quasi-periodic oscillations.

\centerline{\bf Figure Captions}

\noindent {\small {\bf Fig. 1:}
Variation of density (upper-left), radial velocity (upper-right), angular
momentum (lower-left) and Mach number (lower-right) with distance 
in a viscous transonic flow. Different curves are drawn for 
different viscosity parameters. As viscosity is increased, 
the shock is drifted outwards, until it disappears altogether
beyond a critical viscosity parameter. The specific angular
momentum with which matter is injected at the outer boundary is $\lambda=1.8$}                 

\noindent {\small {\bf Fig. 2:}
Same as in Fig. 1 except that the specific angular momentum of the injected 
flow is chosen to be the same as the marginally stable value of $1.8363$.}                 

\noindent {\small {\bf Fig. 3:}
Drifting of the steady shock location in a two dimensional, axisymmetric
accretion flow as viscosity parameter is increased towards the 
critical value. See text for injected parameters.
}                 

\noindent {\small {\bf Fig. 4a:}
Oscillation of the shocks with time when a steady shock is not 
predicted, yet two sonic points exist in a viscous transonic flow.
These solutions provide viable explaination to the quasi-periodic
modulation of the X-rays in blak hole candidates.}

\noindent {\small {\bf Fig. 5:}
Oscillation of the total amount of matter in the accretion disc
with a shock wave. Total particle number $N$  and subsonic particle
number $N_{sub}$ are plotted as functions of time. Parameters are
same as those used to draw Fig. 4(a-b).}                 

\end{document}